\documentclass[oldversion]{aa}  
\usepackage{graphicx}
\usepackage{txfonts}
\usepackage{longtable}
\usepackage{natbib}
\usepackage{ulem}
\bibpunct{(}{)}{;}{a}{}{,}
\def\ewoii{{\rm EW[O{\sc ii}]}}
\def\oii{{\rm [O{\sc ii}]}}
\def\dn{{$D_{n,4000}$}}
\def\hd{{H$\delta$}}
\def\ha{{H$\alpha$}}
\def\hb{{H$\beta$}}
\def\hdf{{H$\delta_{F}$}}

\begin{document}
\title{Star Formation Activities of Galaxies in the Large-Scale Structures at $z=1.2$}
\subtitle{}
\titlerunning{Star Formation Activities at $z=1.2$}
\authorrunning{Tanaka et al.}

\author{M. Tanaka\inst{1}, C. Lidman\inst{2,3}, R. G. Bower\inst{4}, R. Demarco\inst{5}, A. Finoguenov\inst{6,7}, T. Kodama\inst{8}, F. Nakata\inst{9}, P. Rosati\inst{1}
}

\offprints{M. Tanaka}

\institute{European Southern Observatory, Karl-Schwarzschild-Str. 2
	D-85748 Garching bei M\"{u}nchen, Germany
	\email{mtanaka@eso.org}
	\and
	European Southern Observatory, Alonso de Cordova 3107, Casilla 19001, Santiago, Chile 
        \and
        Oskar Klein Center, Roslagstullsbacken 21, 106 91 Stockholm, Sweden
	\and
	Department of Physics, University of Durham, South Road, Durham DH1 3LE, UK
	\and
	Department of Astronomy, Universidad de Concepci\'on. Casilla 160-C, Concepci\'on, Chile
	\and
	Max-Planck-Institut f\"{u}r extraterrestrische Physik, Giessenbachstrasse, D-85748 Garching bei M\"{u}nchen, Germany
        \and
        University of Maryland, Baltimore County, 1000 Hilltop Circle,  Baltimore, MD 21250, USA
	\and
        National Astronomical Observatory of Japan, Mitaka, Tokyo 181-8588, Japan
	\and
	Subaru Telescope, National Astronomical Observatory of Japan, 650 North A'ohoku Place, Hilo, HI 96720, USA
}

\date{Received; accepted }

\abstract{
Recent wide-field imaging observations of the X-ray luminous
cluster RDCS~J1252.9-2927 at z=1.24 uncovered several galaxy groups
that appear to be embedded in filamentary structure extending from
the cluster core.  We make a spectroscopic study of the galaxies in
these groups using GMOS on Gemini-South and FORS2 on VLT with the
aim of determining if these galaxies are physically associated to
the cluster. We find that three groups contain galaxies
at the cluster redshift and that they are probably
bound to the cluster. This is the first confirmation of filamentary
structure as traced by galaxy groups at $z>1$. We then use several
spectral features in the FORS2 spectra to determine the star
formation histories of group galaxies. We find a population of
relatively red star-forming galaxies in the groups that are absent
from the cluster core. While similarly red star forming galaxies can
also be found in the field, the average strength of the \hd\ line
is systematically weaker in group galaxies.
Interestingly, these groups at $z=1.2$ are in an environment in which
the on-going build-up of red sequence is happening.
The unusual line strengths can be explained by star formation that
is heavily obscured by dust.
We hypothesize that galaxy-galaxy interactions, which is more efficient
in the group environment, is the mechanism that drives these dust
obscured star formation. The hypothesis can be tested by obtaining spectral
observations in the near-IR, high resolution imaging observations
and observations in the mid-IR. 
}{}{}{}{}


\maketitle

\section{Introduction}

Over the course of last 13 billion years, the Universe has evolved
from an almost uniform state to the rich diversity of galaxies that we
see locally today. In the local Universe, galaxies in low density
regions (commonly referred to as field galaxies) are typically blue
late-type galaxies.  In sharp contrast to the field, galaxies in high
density regions, such as the cores of rich galaxy clusters, are
dominated by red early-type galaxies. The dichotomy suggests that the
evolution of galaxies is strongly dependent on the environment in
which these galaxies live and that environment must play an essential
role in shaping the Hubble sequence. While a number of processes to
explain this environment-dependent galaxy evolution have been
identified (major and minor mergers, galaxy harassment and ram
pressure stripping, for example) the effectiveness of each of these
processes is still a matter of considerable debate.  A direct way to
unveil the origin of the environmental dependence of galaxy evolution
is to look at galaxies at various redshifts and to quantify how
galaxies change their properties as functions of environment and time.

Galaxy clusters and the regions surrounding them are natural
laboratories for studying galaxy evolution, because of the wide range
of environments - from the dense core to the low density field - that
they contain. An emerging picture from studies of clusters at $z<1$
(e.g., \citealt{dressler97,poggianti99,kodama01,balogh04,demarco05,
  cucciati06,cooper07,koyama07,poggianti08}; see \citealt{tanaka05}
for a more thorough set of references) is that massive red galaxies
were already in place in galaxy clusters by $z\sim1$.
This is in contrast to the field, where a number of studies have shown
that the stellar mass function of red galaxies has evolved strongly at $z<1$
(e.g., \citealt{bell04,bundy06,faber07}).
Although there are
clear signs of galaxy evolution in clusters (e.g., \citealt{cucciati06} and
\citealt{cooper07} found an increasing fraction of blue galaxies at higher
redshifts), the bulk of the stellar mass in clusters at $z<1$ are
contained within these red and apparently dead galaxies.
To understand
when and how these galaxies formed, we now need to move to higher
redshifts. However, the number of galaxy clusters at $z>1$ is still
very few and our understanding of the galaxy population in $z\gtrsim1$
clusters is limited (e.g.,
\citealt{blakeslee03,rosati04,stanford05,nakata05,stanford06,lidman08,hilton09,mei09}).

Most high-$z$ cluster studies are based on broadband photometric
data. However, broadband photometry alone does not allow us to examine
the star formation histories of galaxies in much detail.  Galaxy
spectra, on the other hand, are rich in absorption/emission features
and can be used to provide a more complete picture of stellar
populations and star formation histories. Clusters at $z\gtrsim1$ are
expected to be significantly less evolved than clusters at lower
redshift, so they are an ideal place to understand the processes that
drive galaxy evolution.  Unfortunately, examining $z>1$ clusters with
the same level of detail as that done for $z<1$ clusters is
observationally challenging.  This is not only due to the faintness of
these very distant galaxies, but also because many prominent features
of galaxies at $z>1$ redshift to $>8000\rm\AA$, where
many of the current optical detectors have low sensitivities.
Red sensitive detectors on an 8m telescope, like FORS2 on VLT,
has improved the situation, and it is now feasible to study spectral
properties of galaxies at $z\gtrsim1$ in greater detail.
In this paper, we present detailed analyses of stellar populations
of galaxies around the RDCSJ~1252.9-2927 cluster at $z=1.24$
(hereafter RDCSJ1252).

The cluster was originally discovered with ROSAT and \citet{rosati04}
presented a detailed X-ray analyses based on the deeper Chandra and
XMM-Newton data.  The cluster exhibits a tight cluster red sequence
\citep{blakeslee03,lidman04,mei09}.  \citet{toft04} and
\citet{strazzullo06} derived the $K_s$ band luminosity function of the
cluster galaxies and found that the bright end of the cluster
luminosity function has not significantly changed, except for dimming
due to passive evolution of an old stellar population.  An intensive
spectroscopic follow-up campaign was carried out by \citet{demarco07}.
Within the redshift interval $1.22 < z < 1.25$, \citet{demarco07}
spectroscopically confirmed 38 cluster members and found evidence that
the cluster is undergoing a major merger.  \citet{holden05} performed
a fundamental plane analysis based on the deep spectroscopy of the
four brightest cluster members.  Recently, \citet{gobat08} presented a
sophisticated spectro-photometric analysis of the cluster galaxies and
showed that the cluster galaxies assembled the bulk of their stars
over a short time scale and 0.5 Gyr earlier than field galaxies at similar redshifts.
Also, \citet{rettura08} showed that most of the cluster galaxies have
not actively formed stars for the last $\sim1$ Gyr based on deep
$U$-band imaging.

As part of an on-going survey of distant galaxy clusters
\citep{kodama05}, wide-field multi-band imaging of RDCSJ1252 and the
surrounding region were used to study the large-scale structure
surrounding the cluster. \citet{tanaka07a} reported detecting several
significant clumps extending far out from the cluster core.  To date,
the outskirts of high redshift clusters have not been explored in much
detail. The red sequence in the cluster core extends down to faint
magnitudes, but the red sequence in the surrounding clumps is sharply
truncated at $K_s=22$ \citep{tanaka07a}.  As reported by
\citet{tanaka05}, the red sequence in groups at lower redshifts
extends to fainter magnitudes.  This suggests that some of the
blue galaxies in these clumps at $z=1.24$ will evolve into galaxies
on the red sequence.
In this paper we report on spectroscopic follow-up observations of
these clumps with GMOS on Gemini-South and FORS2 on VLT.
We will use these observations to
discuss physical process(es) driving the build-up of the red sequence.

The paper is structured as follows.  FORS2 and GMOS spectroscopy of
galaxies embedded within the clumps found by \citet{tanaka07a} are
presented in Section 2, followed by a discussion on the nature of
the clumps in Section 3.  In section 4, a comparison between the
spectral properties of these galaxies and those in both the field and the
cluster core will be made.  The results are discussed in section 5 and
the paper is summarized in Section 6.  Unless otherwise stated, we
adopt H$_0=70\rm km\ s^{-1}\ Mpc^{-1}$, $\Omega_{\rm M}=0.3$, and
$\Omega_\Lambda =0.7$.  Magnitudes are on the AB system.  Distances
are reported in the comoving reference frame.

\section{Spectroscopic Observations}

We used GMOS on Gemini-South \citep{hook04} and FORS2 on VLT UT1
\citep{appenzeller98} to obtain
spectroscopic follow-up observations of the structures discovered in
\citet{tanaka07a}. The observations were performed in queue mode, and
six fields were observed. The fields cover the most prominent part
of the structures, as shown in Fig. \ref{fig:field_coverage}.  The
exposure times are listed in Table \ref{tab:data}.

\subsection{GMOS observations\label{sec:data}}

The GMOS observations were taken between January and March 2007 and were
obtained with the R400 grating set to $7800\AA$ and $7900\AA$ to fill
the detector gaps, and the G5325 order sorting filter.
This setting covers the $7000-10000\rm\,\AA$ wavelength
range and results in a resolving power of $R\sim2000$ for a 1\arcsec\
slit. In order to achieve good sky subtraction in a crowded region,
the observations were performed in the nod-and-shuffle mode.

Targets for spectroscopy were selected on the basis of photometric
redshifts. Galaxies with $1.0<z_{phot}<1.3$ and $z$-band magnitudes
brighter than 23 were given the highest priority.  The remaining slits
were filled with targets that had colors consistent with star-forming
galaxies at $z\sim1.2$ on the $R-z$ vs $i-K$ diagram.  This
supplemental selection was taken because our photometric redshifts may
be less accurate for blue galaxies than for red galaxies
\citep{tanaka07a}, and also because the nod-and-shuffle mode allows
one to place slitlets quite densely.

\begin{figure}
\centering
\includegraphics[width=6cm]{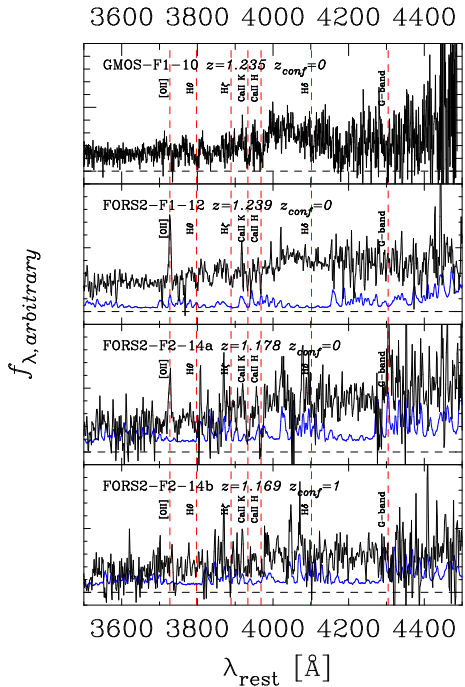}
\caption{
Sample spectra of galaxies in our catalog from GMOS and FORS2.
The dotted line in each panel shows the $1\sigma$ error.
The error is not plotted in the top panel because 
the GMOS pipeline does not produce an error spectrum.
Some of the more prominent spectral features are indicated with
the vertical dashed lines.
The horizontal dashed lines show the zero flux level.
}
\label{fig:spec_eg}
\end{figure}

The data were reduced with the Gemini data reduction package.  The
wavelength and flux calibrated 1D spectra were then inspected with
custom designed software to compute redshifts and assign confidence
flags.  Out of 24 galaxies in field F1, 8 secure redshifts
(representing a 33\% success rate) were obtained.  For field F2, 11
redshifts were obtained from 25 galaxies (44\% success rate), and for
field F3, 14 redshifts were obtained from 31 galaxies (45\% success
rate). The redshifts were obtained by fitting Gaussians to prominent
spectral features such as \oii\ and the Ca{\sc ii}\,H and K lines.
The redshift errors were estimated through Monte-carlo simulations.
We added noise, which was directly estimated from each spectrum, to
each data point of the spectrum (i.e., each point fluctuated by its
noise). The Gaussian fit was then re-performed.  This procedure was
repeated 100,000 times and a 68-percentile interval of the redshift
distribution is quoted as the error.  A confidence flag of 0 indicates
a secure redshift. A confidence flag of 1 indicates a possible
redshift.  Fig. \ref{fig:spec_eg} presents an example from the GMOS
data.

\subsection{FORS2 observations}

The FORS2 observations were taken between April and July 2008 and were
obtained with the $300I$ grism and the OG590 order sorting filter.
This setting covers the $6000-10000\rm\,\AA$ wavelength range and
results in a resolving power of $R\sim500$ for a 1\arcsec\ slit.
Objects were nodded along the slit in order to facilitate accurate
removal of the sky background. Targets for spectroscopy were selected
with photometric redshifts.  Bright galaxies ($z$-band magnitude
brighter than 23) at $1.0<z_{phot}<1.3$ had the highest priority in
the slit assignment.

The data reduction was performed with a custom designed reduction
package.  First, frames were grouped in A-B pairs, where A is one
position of the nod and B is the other. These frames were then
subtracted from each other, after scaling for differences in the sky
background. Cosmic rays were then identified on these sky subtracted
frames and were masked out.  In order to avoid correlations between
pixels, which makes it difficult to estimate flux errors, the 2
dimensional data were not rectified. One dimensional object fluxes and
errors were extracted from each A-B pair.  The fluctuations in the
background are used as a measure of the background noise.  All the 1
dimensional spectra in each slitlet were then combined.
Wavelength shifts caused by instrument flexures
and optical distortions were corrected by tracing the relative
positions of the sky lines.  The wavelengths were then calibrated
using sky lines, and fluxes were calibrated with spectroscopic
standard stars.  The strong telluric absorption at $\sim9300\rm\AA$
was corrected using the spectra of stars that were observed in the
same mask.

All spectra were visually inspected. Redshifts and confidence flags
were assigned in the same way as that done for the GMOS spectra.  Out
of 31 galaxies in field F1, 30 secure redshifts (representing 97\% of
the sample) were obtained.  For field F2, 30 redshifts were obtained
from 36 galaxies (83\% success rate), and for field F3, 13 redshifts
were obtained from 31 galaxies (43\% success rate). The low success
rate in F3 is because only a fraction of the requested total exposure
was obtained (see Table \ref{tab:data}).  The overall success rate of
the FORS2 observations is very high compared to the GMOS observations.
This is largely due to relative sensitivity of the FORS2 CCDs at
$\sim9000\rm\AA$, where the most prominent spectral features of
$z\sim1.2$ galaxies land.  The FORS2 spectra have a high enough signal
for detailed spectral analyses ;
the median signal to noise ratio of the spectra with secure redshifts
is $\sim3$ per \AA at $\rm 3900\AA<\lambda_{rest}<4100\AA$.
We present spectral properties of
galaxies using the FORS2 spectra in Section 4.

\subsection{Combined Spectroscopic Catalog}

The redshifts from the GMOS and FORS2 observations were compiled into
a single spectroscopic catalog. The catalog contains 102 objects with
secure redshifts and 18 objects with possible redshifts. The catalog 
is presented in Table \ref{tab:spec_cat}. One object has a secure
redshift from both FORS2 and GMOS. The difference between the two
measurements is $z_{GMOS}-z_{FORS2}=-0.001$.
To the GMOS and FORS2 catalog, we add objects published in
\citet{demarco07} to create a combined catalog. The combined catalog
covers a wide range of environments at $z=1.24$ -- from the dense
cluster core to the surrounding lower-density field.  In the following
sections, an analysis of the distribution and properties of these
galaxies will be presented. Only objects with secure redshifts are
used in this analysis.

\begin{table}
\caption{Field name and exposure times.}
\label{tab:data}
\centering
\begin{tabular}{l l}
\hline\hline
Field ID   & Exposure Time\\
\hline
GMOS-F1    & 14 $\times$ 30 min \\ 
GMOS-F2    & 14 $\times$ 30 min \\   
GMOS-F3    & 6 $\times$ 30 min  \\  
FORS2-F1   & 12 $\times$ 22 min \\   
FORS2-F2   & 8 $\times$ 22 min  \\  
FORS2-F3   & 2 $\times$ 22 min  \\  
\hline
\end{tabular}
\end{table}

\begin{table*}
\caption{The GMOS and FORS2 spectroscopic catalog, listing 
the object name (ID), the coordinates in J2000, the redshift, the confidence flag
(0 for a secure redshift and 1 for a possible redshift),
and the total magnitudes and their errors.
Magnitudes are on the AB system.
The redshift error does not include wavelength calibration error which is
typically $0.2\rm\AA$.
The magnitude of strongly blended objects is listed with a -1.
{\it The table will appear in its entirety as supplemental material.}}
\label{tab:spec_cat}
\centering
\begin{tiny}
\begin{tabular}{l l l l l l r l r l r l r l r}
\hline\hline
ID & R.A.  & Dec.  & redshift  & flag & $V$ & $\sigma(V)$ &  $R$ & $\sigma(R)$ & $i$ & $\sigma(i)$ & $z$ & $\sigma(z)$ & $K$ & $\sigma(K)$ \\ 
\hline
\vspace{2pt}
   GMOS-F1-4 & 12 53 16.63 & -29 17 11.6 & $1.0706^{+0.0001}_{-0.0001}$ & 0 & 25.40 & 0.12 & 25.46 & 0.12 & 25.03 & 0.12 & 24.69 & 0.20 & 24.63 & 0.81\\ \vspace{2pt}
   GMOS-F1-5 & 12 53 17.44 & -29 17 00.8 & $1.0684^{+0.0001}_{-0.0001}$ & 1 & 24.51 & 0.09 & 24.31 & 0.08 & 23.91 & 0.08 & 23.39 & 0.11 & 22.08 & 0.20\\ \vspace{2pt}
   GMOS-F1-6 & 12 53 09.02 & -29 16 47.6 & $1.0976^{+0.0007}_{-0.0010}$ & 1 & 26.37 & 0.47 & 24.44 & 0.10 & 23.54 & 0.07 & 22.70 & 0.07 & 21.03 & 0.10\\ \vspace{2pt}
   GMOS-F1-9 & 12 53 16.03 & -29 16 17.2 & $1.2452^{+0.0003}_{-0.0002}$ & 1 & 24.23 & 0.20 & 23.57 & 0.10 & 22.71 & 0.07 & 21.77 & 0.07 & 19.87 & 0.07\\ \vspace{2pt}
  GMOS-F1-10 & 12 53 15.43 & -29 16 23.7 & $1.2354^{+0.0004}_{-0.0004}$ & 0 & 24.56 & 0.57 & 22.76 & 0.11 & 21.86 & 0.06 & 20.96 & 0.05 & 19.06 & 0.05\\ \vspace{2pt}
  GMOS-F1-11 & 12 53 14.92 & -29 16 13.4 & $1.2368^{+0.0011}_{-0.0013}$ & 0 & 25.34 & 0.30 & 24.07 & 0.10 & 23.09 & 0.07 & 22.19 & 0.06 & 20.17 & 0.07\\ \vspace{2pt}
  GMOS-F1-12 & 12 53 15.96 & -29 16 29.9 & $1.2839^{+0.0008}_{-0.0007}$ & 1 & 25.49 & 0.28 & 24.08 & 0.08 & 23.67 & 0.09 & 22.86 & 0.09 & 20.98 & 0.11\\ \vspace{2pt}
  GMOS-F1-13 & 12 53 15.28 & -29 16 29.8 & $1.0699^{+0.0006}_{-0.0003}$ & 0 & 26.49 & 0.58 & 23.80 & 0.06 & 23.43 & 0.07 & 22.49 & 0.07 & 20.77 & 0.09\\ \vspace{2pt}
  GMOS-F1-14 & 12 53 16.22 & -29 16 38.4 & $1.0698^{+0.0001}_{-0.0001}$ & 0 & 24.22 & 0.10 & 23.55 & 0.06 & 23.01 & 0.05 & 22.52 & 0.07 & 20.98 & 0.11\\ \vspace{2pt}
  GMOS-F1-21 & 12 53 16.41 & -29 15 44.0 & $1.1797^{+0.0004}_{-0.0005}$ & 1 & 23.32 & 0.04 & 23.14 & 0.03 & 23.19 & 0.05 & 23.25 & 0.11 & 22.29 & 0.25\\
\hline
\end{tabular}
\end{tiny}
\end{table*}

\begin{figure}
\centering
\includegraphics[width=9cm]{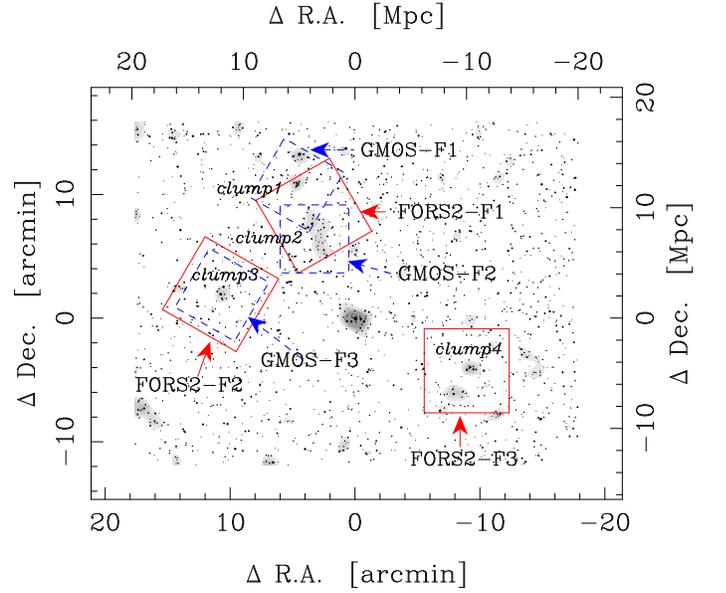}
\caption{
Distribution of photo$-z$ selected galaxies in RDCSJ1252.
The dots show individual galaxies and the shaded regions illustrate the local over-density of such galaxies.
The four clumps reported in \citet{tanaka07a} are shown with the italic labels.
The solid and dashed squares indicate the fields covered by FORS2 and GMOS, respectively.
The top and right-hand axes are in comoving Mpc.
}
\label{fig:field_coverage}
\end{figure}

\section{Large-Scale Structures at $z=1.24$}

\subsection{Spectroscopic confirmation of the structures}

Let us start with the conclusion of this section -- two of the four
clumps identified in \citet{tanaka07a} contain galaxies at the cluster
redshift. The clumps are embedded in a filamentary structure that
extends northwards from the cluster core. In addition, a new clump
is discovered and it lies in a
second structure that extends eastwards from the cluster. Refer to
Fig. \ref{fig:field_coverage} for the location of these clumps.  The
properties of each clump are summarized in Table
\ref{tab:clump_props}.

\begin{figure}[ht]
\centering
\includegraphics[width=9cm]{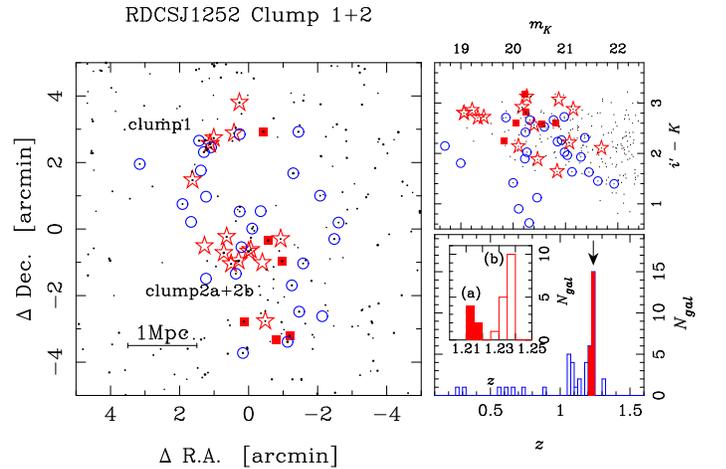}
\caption{
{\it Bottom-right panel:}
The redshift distribution of galaxies. The inset zooms in at the cluster redshift,
which is indicated by the arrow. 
{\it Top-right panel:}
The color magnitude diagram. The filled squares and open stars show galaxies 
at $z\sim1.21$ and $z\sim1.23$, respectively. The dots are galaxies with $1.0<z_{phot}<1.3$.
Open circles are spectroscopically confirmed foreground/background galaxies.
{\it Left panel:}
The spatial distribution of galaxies.  The symbols are the same as those in the top-right panel.
}
\label{fig:spec_zoom_f12}
\end{figure}

\noindent
{\bf Clumps 1 and 2 (Fig. \ref{fig:spec_zoom_f12}) :}
These two clumps lie to the north of the RDCSJ1252 and both contain
concentrations of galaxies close to the cluster redshift.  Clump 1
contains 5 spectroscopically confirmed members. Its redshift is very
close to the cluster redshift ($z=1.237$; \citealt{demarco07}).  Here,
galaxies with $|z-z_{clump}|<0.005$ are defined as clump members.
This window is wide enough to include most members, while it is small
enough to exclude apparent outliers.  The same criterion is used for
the other clumps.  Clump 1 is $12$ Mpc away (transverse distance) from
the cluster core .

Clump 2, which lies in between clump 1 and the cluster core, is also
confirmed to be real.  Interestingly, the redshift distribution of
clump 2 is bimodal as indicated in Fig. \ref{fig:spec_zoom_f12} -- one
peak at $z=1.214$ and the other at $z=1.234$.  The line of sight
distance between the two peaks is $\sim40$ Mpc.  They are called
clump 2a and 2b in Table \ref{tab:clump_props}, respectively.  Clump 2a
has 6 spectroscopically confirmed members, however, the galaxies are
not spatially concentrated very strongly.  On the other hand, clump 2b shows
a clear spatial concentration of galaxies at the cluster redshift.

Extended X-ray emission from clumps 1 and 2 was reported in
\citet{tanaka07a}. Clump 2b is most likely the counterpart of the
detected X-ray emission in that area because clump 2a has a looser
spatial distribution.  As we will show in the next subsection, clump 2a
is dynamically independent from the RDCSJ1252 cluster. The detection
of X-ray emission from both clump 1 and clump 2b support the notion
that both clumps are bound systems.

\begin{figure}[ht]
\centering
\includegraphics[width=9cm]{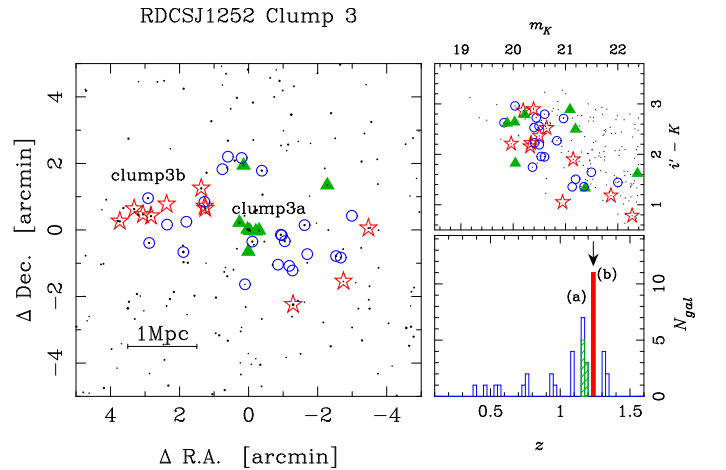}
\caption{
Same as Fig. \ref{fig:spec_zoom_f12}, but for clumps 3a (hatched histogram and filled triangles) 
and 3b (filled histogram and open stars).
}
\label{fig:spec_zoom_f3}
\end{figure}

\noindent
{\bf Clump 3 (Fig. \ref{fig:spec_zoom_f3}) :}
This clump is located 10\,Mpc eastwards from the cluster core.  The
spectroscopy shows a strong concentration at $z=1.175$, which is more
than 100 Mpc away from the cluster along the line of sight.  We call
this system clump 3a.  Although it is likely that clump 3a is a
physically bound system with its strong redshift and spatial
concentrations, this clump is dynamically independent from the cluster
at $z=1.237$.  This clump is marginally detected in X-rays at the
$2.4\sigma$ level \citep{tanaka07a}.

It is interesting to note that there is a second strong redshift peak at the
cluster redshift in this area. A clump of galaxies 2\,Mpc eastwards of
clump 3a is responsible for this peak. This clump is labeled clump 3b
and is likely to be a physically bound system given the clear
redshift and spatial concentration.  Clump 3b indicates that
there is a structure extending eastwards from the RDCSJ1252 cluster.

\begin{figure}[ht]
\centering
\includegraphics[width=9cm]{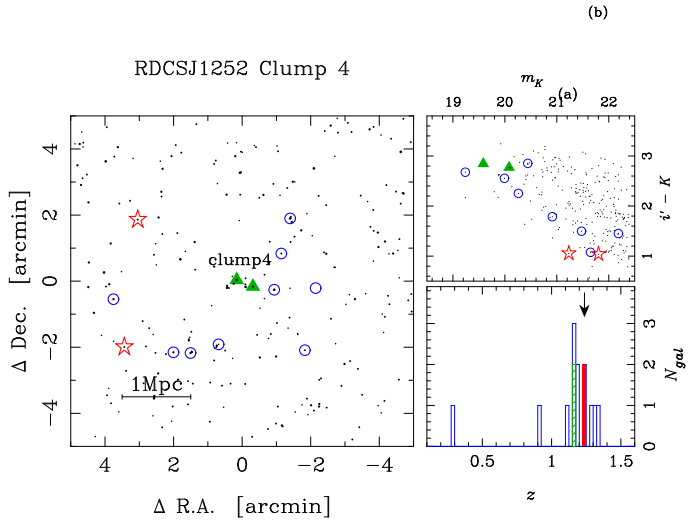}
\caption{
As for Fig. \ref{fig:spec_zoom_f12}, but for clump 4.
}
\label{fig:spec_zoom_f4}
\end{figure}

\noindent
{\bf Clump 4 (Fig. \ref{fig:spec_zoom_f4}) :}
Only 44min of integration was obtained in this field, which resulted
in relatively few redshifts. Two galaxies at the center of the
over-density have $z\sim1.17$ and it is thus likely that clump 4
is a foreground system.  We observe two other galaxies at $z=1.15$
and another two at $z=1.19$ in the same field; however, they land
outside the clump. Overall, there are too few redshifts to decide
if clump 4 is a bound system.

\begin{figure*}
\centering
\includegraphics[width=14cm]{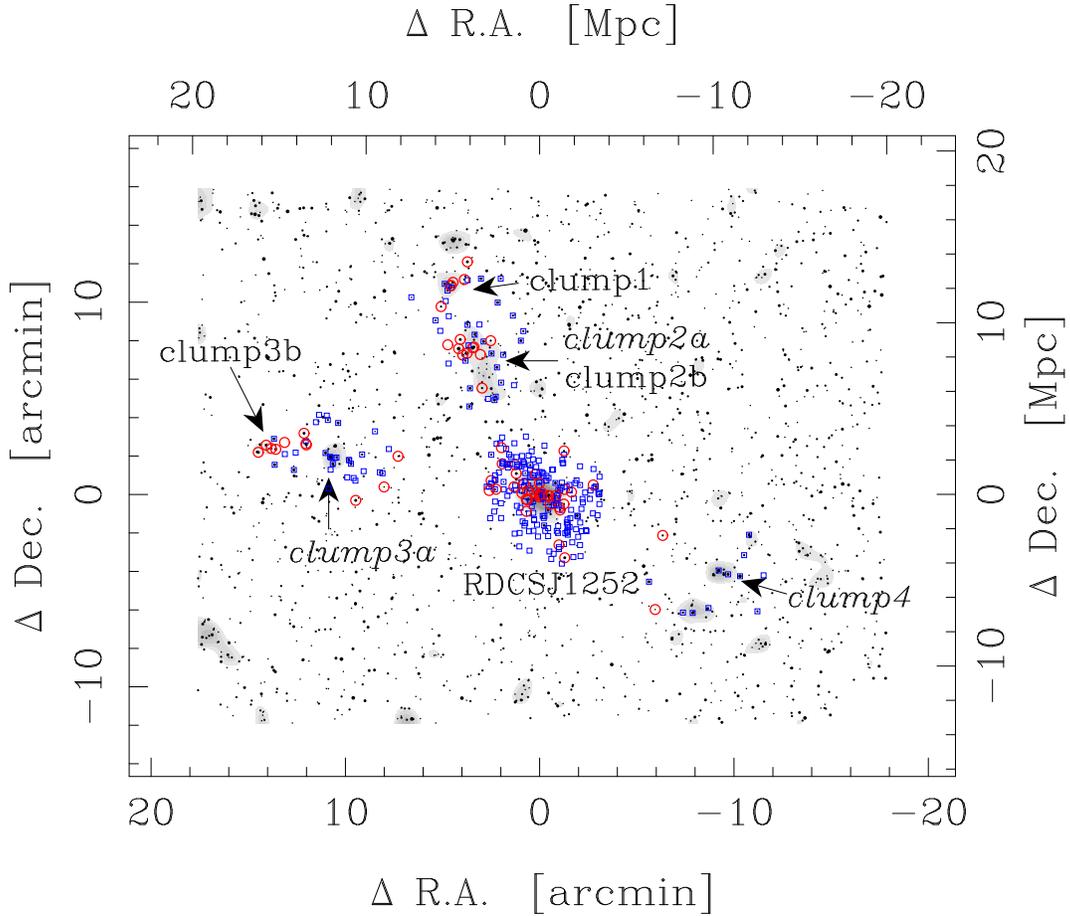}
\caption{
The large-scale structure surrounding RDCSJ1252. 
Dots represent photo$-z$ selected galaxies and the shaded regions show the local over-density
of these galaxies. The spectroscopically confirmed bound systems are indicated with labels
(the italic labels show the foreground systems).
Circles are galaxies at $1.22<z<1.25$ and squares are
those outside of this redshift range.
Note that the RDCSJ1252 cluster is at $z=1.237$ \citep{demarco07}.
}
\label{fig:lss_spec}
\end{figure*}

The large-scale structure surrounding RDCSJ1252 is illustrated in
Fig. \ref{fig:lss_spec}.  Galaxies from \citet{demarco07} are also
plotted.  At least two structures are detected: a filament extending
northwards from the center of the cluster and hosting clumps 1 and 2b
and a group lying eastwards of the cluster and hosting clump 3b.

\subsection{Dynamical analysis}

To quantify the relationship between the clumps and the
cluster, we use the Newtonian energy integral formalism
\citep{beers82,hughes95,lubin98} to compute the probabilities that the
clumps are bound to RDCSJ1252.  We introduce the
position angle $\phi_d$ between a clump and the cluster relative to
the line-of-sight.  The linear distance between them is then given by
$d=D_p/\sin\phi_d$, where $D_p$ is the transverse distance.
Similarly, the relative space velocity between them is represented as
$v=v_r/\cos\phi_v$, where $v_r$ is the line-of-sight velocity
difference and $\phi_v$ is the angle in velocity space relative to
the line-of-sight.  The condition that a clump and the cluster are
bound is then expressed as

\begin{equation}\label{eq:bound_prob}
v_r^2-(2GM/D_p)\sin\phi_d\cos^2\phi_v<0,
\end{equation}

\noindent
where $M$ is the total mass of the system.  The probability that a
clump is dynamically bound to the cluster is equivalent to estimating
the fraction of the solid angle ($\phi_d$ and $\phi_v$) that satisfies
Eq.~\ref{eq:bound_prob}.

Assuming that RDCSJ1252 is virialized, we derive a cluster mass of
$\rm M_{200}=3.7^{+1.0}_{-1.1}\times10^{14}\rm M_\odot$ from the
galaxy velocity dispersion.  Within the errors, this estimate is
consistent with the one derived from X-ray observations
($2.7\times10^{14}\rm M_\odot$; \citealt{rosati04}).
\citet{demarco07} reported that the cluster might consist of two
merging clusters and that the total mass of the two clusters is
$2.7\times10^{14}\rm M_\odot$, consistent with the X-ray estimate.
Cluster masses can be estimated in a number of ways, but each method has
its own systematic uncertainties (e.g., masses from velocity
dispersions can be affected by substructure). Considering these
estimates, we use a mass of $3^{+1}_{-1}\times10^{14}\rm M_\odot$ 
for deriving the bound probabilities. The probabilities are
summarized in Table \ref{tab:clump_props}.

\noindent
{\bf Clump 1 :}
Clump 1, with a bound probability of $\sim 90\%$, is probably bound to RDCSJ1252,
although its transverse separation is 12\,Mpc.

\noindent
{\bf Clump 2 :}
Clump 2b, with a bound probability of $\sim50\%$, may also be bound
to the cluster. Therefore, the filamentary structure extending
northwards of the cluster core is within the gravitational reach of
the cluster. Clump 2a is probably not a bound system in itself given
the loose spatial distribution of the galaxies within this clump.
Assuming that it has a similar mass and transverse distance as clump 2b
(note that small changes to the clump mass and transverse distance
have little effects on the bound probability - the mass of the central
cluster and line-of-sight velocity difference are the key
parameters.), the bound probability is 0, suggesting that the structure
hosting clump 2a is unrelated to RDCSJ1252.

\noindent
{\bf Clump 3 :}
Clump 3a is a foreground system at $z=1.17$ and is not bound to the
cluster ($0\%$ bound probability).  Clump 3b provides evidence for structure extending
eastwards of the cluster. It is bound to the cluster with a $\sim50\%$
probability, a relatively high probability considering it has a
transverse separation of 15\,Mpc from the cluster.  The cluster itself
is elongated in the East-West direction \citep{demarco07}, thus
suggesting that it may have experienced a merger with a group of
galaxies that fell along the direction leading to clump 3b.

\noindent
{\bf Clump 4 :}
There are not enough redshifts to estimate the mass of clump 4. Using
the mass for clump 3a as an upper limit (note that clump 3a is
marginally detected in X-rays whereas clump 4 is not), it is unlikely
(bound probability of 0\%) that clump 4 is bound to the cluster.

To sum up, some of the spectroscopically confirmed structures are
probably bound to RDCSJ1252.  There appears to be two filaments
leading to the cluster: a very clear one extending towards the north
hosting clumps 1 and 2b, and another one extending towards the east
hosting clump 3b.  The structure containing clump 2a at $z=1.21$ and
extending to the north is probably not associated to the central
cluster as suggested from its bound probability.  The other two
clumps, clump 3a and clump 4, are dynamically independent of the
cluster.

This is the first secure confirmation of filamentary structure at
$z>1$ as traced by groups of galaxies.  Such groups enable us to
examine how environment affects the evolution of galaxies at high
redshifts.  We take this opportunity and look into the spectral
properties of these galaxies in the next section using the high
quality FORS2 spectra described in Sec.~\ref{sec:data}.

\begin{table*}
\caption{
Properties of the clumps. The second column shows the number of spectroscopically 
confirmed members in each clump. Redshifts (3rd column) and velocity dispersions 
(4th column) are estimated with the biweight estimator \citep{beers90}.
The redshift errors are estimated by bootstrapping,
taking into account individual redshift errors.
The 5th and 6th columns show the Virial radii and masses \citep{carlberg97}.
The last column shows the probability that each clump is dynamically bound to
RDCSJ1252. The galaxies in clump2a are not spatially concentrated,
suggesting that this clump probably not a bound system. There are only two members in clump 
4 and the redshift is an average redshift of these two galaxies. The properties 
of RDCSJ1252 are added for reference. See text for details.
}
\label{tab:clump_props}
\centering
\begin{tabular}{lllllll}
\hline\hline
           & $N_{spec. member}$ & $z$                  & $\sigma$  [$\rm km s^{-1}$] & $R_{200}$ [Mpc]         & $M_{200}$ [$M_\odot$]            & $P_{bound}$\\
\hline\vspace{2pt}
RDCSJ1252 cluster & 38 & $1.2370_{-0.0011}^{+0.0009}$ & $754_{ -89}^{ +55}$         & $0.93_{-0.11}^{+0.07}$  & $3.7_{-1.1}^{+0.9}\times10^{14}$ & ---\\\vspace{2pt}
Clump1    & 5         & $1.2365_{-0.0015}^{+0.0002}$  & $ 44_{  -5}^{+271}$         & $0.05_{-0.01}^{+0.33}$  & $7.6_{-2.5}^{+270}\times10^{11}$ & $86^{ +7}_{-36}$\\\vspace{2pt}
Clump2a   & 6         & $1.2135_{-0.0013}^{+0.0010}$  & $287_{-159}^{ +36}$         & $0.36_{-0.20}^{+0.04}$  & $2.1_{-1.9}^{+0.9}\times10^{13}$ & $ 0^{ +0}_{ -0}$\\\vspace{2pt}
Clump2b   & 10        & $1.2343_{-0.0013}^{+0.0012}$  & $517_{-164}^{ +97}$         & $0.64_{-0.20}^{+0.12}$  & $1.2_{-0.8}^{+0.8}\times10^{14}$ & $51^{+28}_{-28}$\\\vspace{2pt}
Clump3a   & 8         & $1.1748_{-0.0019}^{+0.0015}$  & $435_{-200}^{ +32}$         & $0.55_{-0.25}^{+0.04}$  & $7.3_{-6.2}^{+1.7}\times10^{13}$ & $ 0^{ +0}_{ -0}$\\\vspace{2pt}
Clump3b   & 8         & $1.2350_{-0.0008}^{+0.0005}$  & $235_{ -94}^{ +41}$         & $0.29_{-0.12}^{+0.05}$  & $1.1_{-0.9}^{+0.7}\times10^{13}$ & $47^{+28}_{-27}$\\\vspace{2pt}
Clump4    & 2         & $1.1745$                      & ---                         & ---                     & ---                              & $0^{+0}_{-0}$\\
\hline
\end{tabular}
\end{table*}

\section{Spectral Analyses}

Following the confirmation of the large-scale structure surrounding
RDCSJ1252, we now use the FORS2 spectra presented here and published
elsewhere \citep{holden05, vanderwel05, demarco07} to examine the
spectral properties of galaxies as a function of environment.  We
define three environments -- cluster, group, and field.  Cluster
galaxies are spectroscopically confirmed members within 1\,Mpc
(physical) of the cluster center \citep{demarco07}. The group galaxies
are defined as those that are within the clumps confirmed in this
paper, except for clump2a as it is probably not dynamically bound.
Clump 3a is a dynamically independent foreground group, but we include
it to gain statistics given that it and RDCSJ1252 have similar
redshifts ($\Delta z=-0.06$).  Note that our results remain
unchanged if we exclude it.  The other galaxies at $1.15<z<1.25$
(e.g., galaxies in the clump2a) are regarded as field galaxies.
In the following, we first
study the star formation activity in these galaxies
by examining \oii\ equivalent widths, strengths of the 4000\AA\ break, and
broad-band $i-K_s$ colors.  We then examine \hd\ equivalent widths.

A fraction of the observed \oii\ emissions might arise from AGN activity
rather than from star formation.
To quantify the AGN contamination, we search for point sources
in the 0.5--2 keV band data of the XMM observations reported in \citet{tanaka07a}.
Within the area covered by XMM, we have recovered all point sources with
$L_X$ exceeding $7\times10^{42}$ $ergs\ s^{-1}$ in the rest frame
0.5--2 keV band assuming a power law of $\Gamma=1.8$ and
the redshift of the cluster, therefore sampling well the typical AGN luminosities.
We find that none of the galaxies we discuss below are detected in X-ray,
giving AGN fractions of $0^{+0.029}_{-0}$, $0^{+0.035}_{-0}$,  $0^{+0.084}_{-0}$,
in cluster, group, and field environments, respectively.
The errors are Poisson errors.
These fractions are fully consistent with those from the COSMOS survey \citep{silverman09},
although X-ray flux limits and stellar mass limits are different and comparisons
are not straightforward.
We regard the observed \oii\ emissions arise from star forming regions, not from AGNs.
Note that \citet{yan06} found that $\sim30$ \% of red galaxies at $z=0$
are LINERs, but such weak AGNs are unlikely to affect our results significantly.


\subsection{Star Formation Activity as a Function of Environment}

\begin{figure}
\centering
\includegraphics[width=9cm]{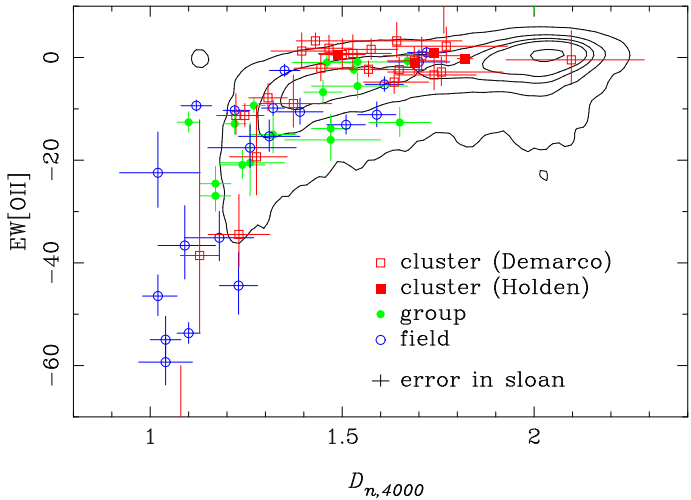}
\caption{
\ewoii\ plotted against \dn\ for galaxies in the cluster, the groups and
the field. The symbols are defined in the plot. The distribution 
for galaxies in SDSS is plotted as contours for comparison. The 
contours enclose 5, 25, 50, 75 and 95 percent of the galaxy population.
The typical error in the SDSS data is shown in the plot.
}
\label{fig:oii_d4000}
\end{figure}

The \oii\ equivalent width (\ewoii) is plotted against the strength
of the 4000\AA\ break (\dn) in Fig. \ref{fig:oii_d4000}.  The \ewoii\
is estimated following \citet{fisher98}. For the 4000\AA\ break, the
definition specified in \citet{balogh99} is used. This definition uses
narrow wavelength windows to measure the break, which minimizes the
effects of the strong telluric absorption feature at $\sim9300\rm\AA$.
Although we remove the absorption from this region
(see Sect. 2.2), we avoid this
region in the analysis in order to minimize systematic errors.
Measurement errors of these spectral indices are estimated via
monte-carlo simulations.  For the cluster data, we estimate the average
noise around the spectral features.  For the group and filament data,
we have a noise spectrum for each object.  The noise is randomly added
to each data point of the spectra and we repeated the measurements
100,000 times.  A 68-percentile interval is taken as an error.

Local galaxies from \citet{tanaka07b} are plotted for comparison.
They are from the  Sloan Digital Sky Survey (SDSS; \citealt{york00,strauss02})
and have $M_V<M_V^*+1$ at $0.005<z<0.065$.
Our spectroscopy reaches a $z$-band magnitude of $\sim23$,
which roughly corresponds to $m_z^*+1$ at $z=1.2$ 
for red galaxies with $z-Ks\sim2$ \citep{toft04,strazzullo06}.
The magnitude cut applied to the SDSS data is relative to $M_V^*$  at $z=0$.
We thus sample the similar luminosity ranges both at $z=0$ and 1.2
relative to the evolving characteristic magnitudes measured at each redshift.
We use the SDSS data only to illustrate the range of spectral indices
that $z=0$ galaxies have.

Most of the spectra in \citet{demarco07} were taken before FORS2 was
upgraded with the red sensitive MIT/LL detectors, so the flux
calibration beyond $9000\AA$ is not sufficiently precise enough for
computing \dn. In order to use these data, a posteriori corrections to
the calculated values of \dn\ were applied using the $i-z$ colors.
The corrections applied are typically $\Delta D_{n,4000}\sim0.1$.
The validity of applying this correction was verified using the
high-signal to noise spectra in \citet{holden05}. There,
$D_{n,4000,calib}-D_{n,4000,Holden}\sim-0.1$. We adopt 0.1 as our
systematic error in estimating \dn\ for the spectra from \citet{demarco07}.
The impact on computing \ewoii\ is negligible as it is measured in a
narrow wavelength window.

Galaxies at $z\sim1.2$, independent of environment, show smaller \dn\
than local galaxies.  For example, galaxies at $z=0$ form a locus of
red galaxies at $D_{n,4000}\sim2$, while red galaxies at $z\sim1.2$
have $D_{n,4000}\sim 1.6$.  This is consistent with passive evolution
of red galaxies from $z\sim 1.2$ to $z\sim 0$.  Blue galaxies at
$z\sim1.2$ are bluer than local blue galaxies, suggesting more active
star formation at $z\sim1.2$ than $z=0$ \citep{blanton06b}.  Overall,
all types of galaxies at $z\sim1.2$ are bluer than local galaxies.

At $z\sim1.2$, there are galaxies in the field and in the groups that
have both red colors ($D_{n,4000}\sim1.6$) and significant \oii\
emission (\ewoii$<-10$\AA).  The fractions of red star-forming
galaxies, defined here as $1.4<D_{n,4000}<1.7$ and \ewoii$<-10$\AA, in
groups and in the field are $0.33\pm0.22$ and $0.40\pm0.33$,
respectively.  Such red star-forming galaxies are not observed in the
cluster. In clusters at slightly lower redshifts
(e.g.,\citealt{geach06,marcillac07,koyama08}) it has been shown that
dusty, star-forming galaxies preferentially reside in low to
medium density regions.
For reference, the fraction of red star forming galaxies in the local
Universe (defined here as $1.8<D_{n,4000}<2.2$ and \ewoii$<-10$\AA) is
$0.06\pm0.01$.

\begin{figure}
\centering
\includegraphics[width=6cm]{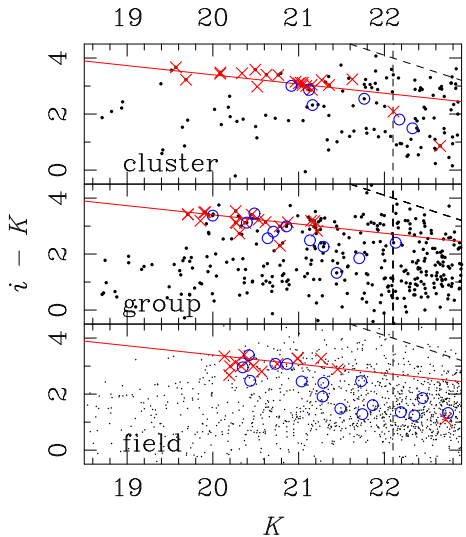}
\caption{
$i-K$ plotted against $K$ for galaxies in the cluster, group and
field environments. The open circles (crosses) represent galaxies
with (without) significant \oii\ emission (EW[O{\sc ii}]$<-10\rm\AA$).
The points show all objects within 1\arcmin\
from the center of each clump. The dashed lines show $5\sigma$ detection limits.
The solid straight lines are fits to the cluster red sequence.
}
\label{fig:cmd_spec}
\end{figure}

We further illustrate the characteristics of these red star-forming
galaxies with a color-magnitude plot, as shown in
Fig. \ref{fig:cmd_spec}.  The $i$ and $K$ band photometry is from
Suprime-Cam and WFCAM, respectively \citep{tanaka07a}.
The cluster is dominated by red galaxies without
strong \oii\ emission.  A tight red sequence in groups is also formed
by galaxies without \oii, but there are some \oii\ emitting galaxies
on the red sequence.  The field galaxies show a significant fraction
of red star-forming galaxies even on the red sequence, although the
red sequence is broader than it is for the groups and the cluster.
We note that the slightly wider redshift range adopted for the field sample
does not strongly contribute to the observed large scatter.
At $1.15<z<1.25$, the $k$-correction for the $i-K_s$ color is
only $\sim0.1$ mag for passively evolving galaxies, which is small
compared to the observed color spread of the field galaxies
($2.5<i-K_s<3.4$ at $K_s\sim20.4$).
We perform a linear fit to the cluster red sequence, shown
as the solid lines in Fig. \ref{fig:cmd_spec}, and find that the
fractions of \oii\ emitters (\ewoii$<-10$\AA) within $\Delta
|i-K|<0.5$ mag from the red sequence are $0.09\pm0.10$, $0.26\pm0.14$,
and $0.27\pm0.16$ in the cluster, group and field environments,
respectively.  Although the errors are large, there are more
star-forming galaxies on the red sequence in group and field
environments.  We will discuss these galaxies in detail later.

The figure shows that we typically sample red galaxies with $20<K<21.5$
and blue galaxies down to $K\sim22.5$ in all the environments.
The cluster galaxies are from \citet{demarco07}, but their magnitude
and color distributions are not strongly different from
those of the group and field galaxies.
Therefore, the spectral differences we observe are unlikely
due to sample biases.
Note that there are more massive galaxies in higher density regions,
but it is due to the environmental dependence of the stellar mass function.
We discuss the stellar mass dependence of the spectral indices later.

\subsection{Detailed Star Formation History of Galaxies}

We now take a deeper look into the spectral properties of the galaxies
in the groups and field surrounding RDCSJ1252.  Among many spectral
features that are sensitive to star formation, the \hd\ absorption is
particularly interesting because it is sensitive to star formation
rates about 0.1--1Gyr prior to the observed epoch.  It is noteworthy
that \hd\ at $z=1.24$ lands in a region that is free of bright night
sky lines. The gap is one of the reddest OH windows in the optical
regime, and $z=1.24$ is probably the highest redshift at which \hd\ is
still accessible with optical spectrographs.  Our FORS2 spectra are
good enough to measure \hd\ line strengths.  Unfortunately, most of
the cluster spectra from \citet{demarco07} were taken with the
original FORS2 CCD, which was not very sensitive beyond $9000\AA$.
The H$\delta$ absorption is around $\sim9200\rm\AA$ in the observed
frame and these cluster spectra do not have a sufficient signal there.
We instead use the spectra from \citet{holden05} with the caveat that
they are the brightest cluster members, so they sample a luminosity
range that is not available to galaxies in groups and in the field.

\begin{figure}
\centering
\includegraphics[width=9cm]{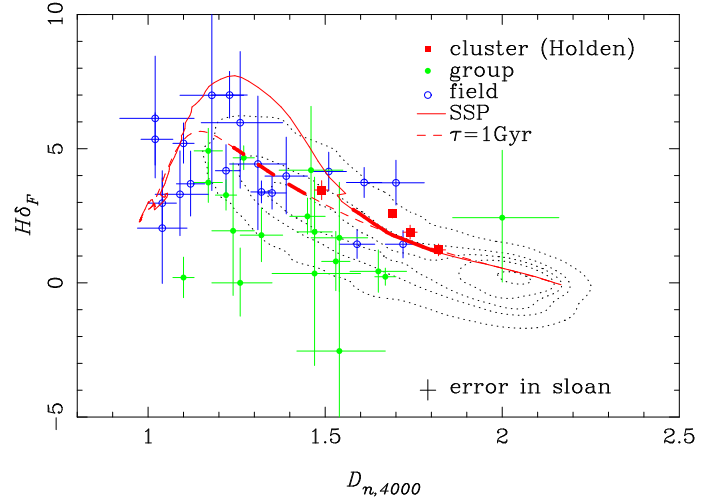}
\caption{
\hdf\ plotted against \dn\ for galaxies in the cluster (red squares), the groups (filled circles)
and the field (open circles). The contours show the distribution of galaxies from the SDSS
for comparison.  The contours enclose 5, 25, 50, 75 and 95 percent of the galaxy population.
The typical error for the SDSS data is shown in the plot.
The solid and dashed lines trace the path of the passively evolving and $\tau=1$ Gyr models, respectively.
The models start from the left (0 Gyr) and evolve to the right (13 Gyr).
The thick part of the lines indicate the range of formation redshifts ($2<z_f<5$) for
galaxies at $z=1.2$.
}
\label{fig:hd_d4000}
\end{figure}

We use the definition of \hdf\ described in \citet{worthey97}, which
uses a narrower spectral window than H$\delta_A$.  The \hd\ line at
$z=1.24$ is close to a strong telluric absorption feature, which lands
at $\sim9300\rm\AA$, and we prefer the narrow window to minimize
effects from this feature, although we do correct all spectra for it.
The contribution to \hd\ from nebular emission was corrected using the
strength of the \oii\ emission line.  The \oii\ flux was first
translated into an H$\alpha$ flux assuming the local relation from
\citet{kennicutt98}, which was then translated into an \hd\ flux
assuming $A_{H\alpha}=1$ mag.~of extinction and the extinction law
from \citet{cardelli89}.  To be specific, we used the relation
H$\delta$=0.09\oii.  In Fig.~\ref{fig:hd_d4000} we plot H$\delta$
against \dn. Local galaxies from the Sloan Digital Sky Survey
are plotted for comparison.

The most striking trend is that group galaxies tend to show lower
\hdf\ than cluster and field galaxies.  This trend is particularly
clear for blue galaxies, e.g., $D_{n,4000}<1.4$.  Blue galaxies in
groups have smaller \hdf\ than blue galaxies in the field by
$\sim3\AA$.  Local galaxies form a clear sequence, and field galaxies
at $z=1.2$ are on the same sequence with an extension to smaller \dn.
On the other hand, group galaxies occupy only the bottom half of the
sequence, suggesting that they have systematically smaller \hdf.  The
cluster galaxies occupy the upper half, but, as noted above, one
should be careful comparing these galaxies to the others because they
are the brightest cluster members. The group and field galaxies are
from the same observing run and the data reduction and spectral
analysis were performed in exactly the same way.  It seems
inconceivable that the systematic offset in \hdf\ is due to
observation biases. The trend seems to be real. Recall that
\citet{tanaka07a} report that the faint end of the red sequence in
these groups was truncated. Some of the group galaxies must be
changing their properties, and the weak \hd\ line provides us with a
clue to identifying the main physical process behind the build-up of
the red sequence.

In order to better understand the significance of the difference
between field and group galaxies, we have added predictions from
models to the plot. \citet{bruzual03} models are constructed with two
star formation histories; a single burst model (denoted as SSP) and a
$\tau$ model with $\tau=1$\,Gyr.  We assume solar metallicity, a
Salpeter initial mass function and no dust extinction (see
\citealt{bruzual03} for details).  A sharp truncation of star
formation makes \hd\ very strong $\sim0.5$ Gyr after the burst, while
a gradual truncation does not trigger such an enhancement in \hd\
(e.g., \citealt{couch87,poggianti99}).  Very blue galaxies in the
field have very strong \hd\ absorptions, and they are consistent with
an early phase of the single burst model.  They may be undergoing
intense starbursts.  Such very blue galaxies are rare at $z=0$ as seen
from the distribution of the SDSS galaxies.  The models do not reproduce
the observed \hdf\ in group galaxies. Many group galaxies have weaker
\hdf\ than the $\tau=1$ Gyr model.

We further quantify star formation histories by measuring specific
star formation rates (SSFR). The SSFR is defined as $SSFR=SFR / M_*
\rm\ [yr^{-1}]$, where $M_*$ is the stellar mass.  Stellar masses are
derived from the $K$ band photometry taken with WFCAM
\citep{tanaka07a}.  We use \citet{bruzual03} models with different
values for $\tau$ to estimate the stellar mass from the $K$ band
luminosity and the $i-K$ color. We assume solar-metallicity, a
Salpeter initial mass function and no dust extinction. SFRs are
estimated from the \oii\ emission line \citep{kennicutt98}.  We apply
a correction factor of $\sim30\%$ for slit losses. We assume that all
of our objects are point sources and estimate the slit losses from stars
that were observed at the same time as the group and field galaxies.

\begin{figure}
\centering
\includegraphics[width=9cm]{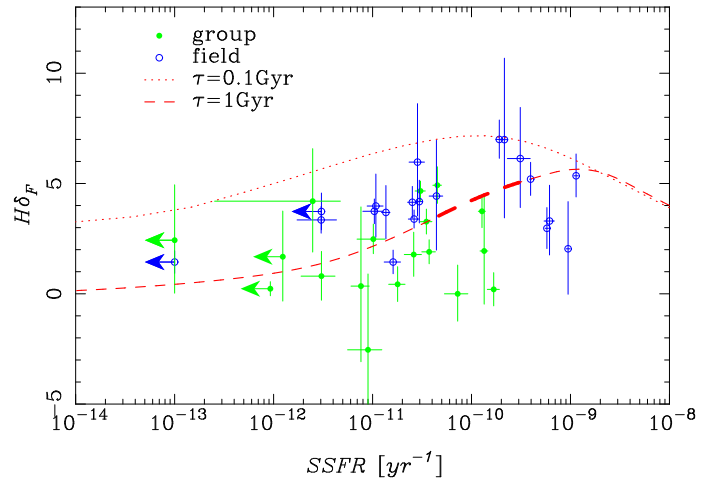}
\caption{
\hdf\ plotted against specific SFR for galaxies in groups (filled circles) and the field (open
circles). The dotted and dashed lines trace the path of the 
$\tau=0.1$ and $\tau=1$\,Gyr models, respectively. The models start from the right and evolve to 
the left. The thick part of the lines indicate the range of formation redshifts $2<z_f<5$ for
galaxies at $z=1.2$.
For $\tau=0.1$\,Gyr model, this redshift range is outside the plotted range.
}
\label{fig:hd_ssfr}
\end{figure}

We plot \hdf\ against SSFR in Fig. \ref{fig:hd_ssfr} along with two
models representing different star formation timescales, a slowly
decaying one with $\tau=1$\,Gyr and a rapidly decaying one with
$\tau=0.1$\,Gyr.  We cannot draw SSP model in the figure because SSFR
goes to zero immediately after the burst.  Instead we plot a
$\tau=0.1$ Gyr model to show the effect of different star formation
time scales.  Note that the SSFR shown in this figure is primarily
sensitive to the current star formation activity, while \dn\ is
sensitive to the integrated star formation history.

Galaxies with different star formation time scales occupy different
parts of the plot.  A short time scale model will enhance \hd\ at early
times, and the absorption remains relatively strong at small SSFRs.
On the other hand, a long time scale model keeps \hd\ weaker over
almost all the plotted SSFR range.  The position of a galaxy on the \hd-\dn\
plot can be used to estimate the timescale on which galaxies stop
forming stars based on the assumption of the exponentially decaying
star formation rate.

\begin{figure}
\centering
\includegraphics[width=9cm]{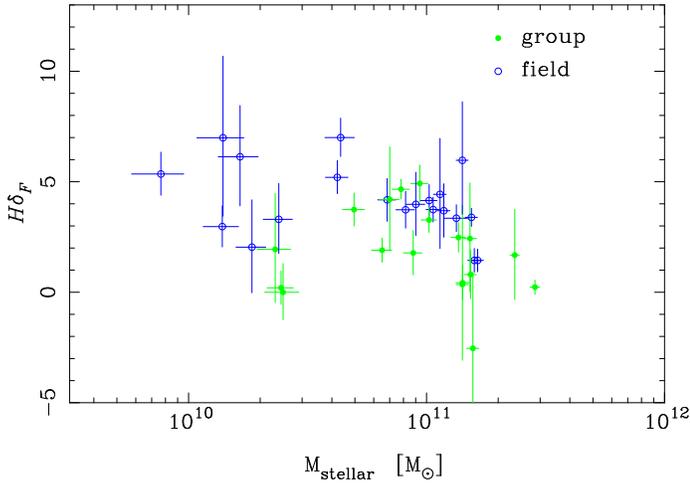}
\caption{
\hdf\ plotted against stellar mass for galaxies in groups (filled circles) and
field (open circles).
}
\label{fig:hd_mstar}
\end{figure}

The distribution of field galaxies in Fig.~\ref{fig:hd_ssfr} (i.e.,
large SSFR) suggests that these galaxies have relatively short star
formation time scales.  They may be currently experiencing very active
star formation.  As seen in Fig~\ref{fig:hd_d4000}, a systematic
offset in \hdf\ for galaxies in groups, in the sense that it is weaker
in groups, is also seen in Fig.~\ref{fig:hd_ssfr}.
The trend is not due to a strong mass dependency of
galaxy properties (e.g., \citealt{vanderwel05}).
As a sanity check, we plot \hdf\ against stellar mass in Fig. \ref{fig:hd_mstar}.
Group galaxies have \hdf\ systematically weaker than field galaxies.
Using galaxies with $>5\times10^{10}\rm M_\odot$, where we are
reasonably complete, the median \hdf\ indices are 1.8 and 3.7 for
group and field galaxies, respectively.
The Kolmogorov-Smirnov test suggests that the two populations are
drawn from the same parent population at $<1$\% confidence level.  
The trend is therefore an environmental variation.
We highlight again that the two samples are from
the same observing run, and we reduced and analyzed them in exactly the same way.
The difference is unlikely due to observation or selection biases.

The \hdf\ index can place constraints on
the time scale on which galaxies quench their star formation
activities as shown by the models in Figs. \ref{fig:hd_d4000} and
\ref{fig:hd_ssfr}.  There are several physical processes that could
lead to the truncation of star formation.  The time scale over which this
occurs (the quenching time scale) is an important observational
constraint, as it may help us understand the dominant process.  In the
next section, we will further discuss the trends we observe here and
address implications of our results for galaxy evolution.

\section{Discussions}

We carried out the spectroscopic follow-up observations of the
surrounding regions of the RDCSJ1252 cluster.  We confirmed some of
the photometrically identified structures and there seems to be at least two
structures extending from the cluster.
The large-scale structures at $z=1.24$ give us
a unique opportunity to look into star formation histories of
galaxies as a function of environment.  In this section, we extend
discussions in the last section and address roles of environments on
galaxy evolution.

Our primary findings in the previous section are summarized as follows.

\begin{enumerate}
\item[1] There is a population of star-forming galaxies with red
  colors in groups and the field at $z\sim1.2$. The core of RDCS1252
  has very few galaxies of this type.

\item[2] Galaxies in groups have systematically weaker \hd\ lines than
field galaxies.
\end{enumerate}

A natural explanation for the first finding - the red colors of the
star-forming galaxies - is that 
these galaxies are reddened by dust. Recent mid-infrared
observations of clusters have shown that dusty star forming galaxies
are found in the outskirts of galaxy clusters. They tend to avoid the
cluster core (e.g., \citealt{geach06,marcillac07,koyama08}).
The same trend has been observed at $z\sim0.2$ \citep{gallazzi09,wolf09},
although it does not necessarily hold in merging clusters \citep{haines09}\footnote{
We note that \citet{marcillac08} showed that the distribution of
stellar mass selected LIRGs and ULIRGs does not depend on environment.
The difference might be due to large-scale structures.
\citet{marcillac08} studied a blank field, while the other authors
studied cluster fields, where structures are prominent.
Isolated groups and those in rich structures might have different properties
(see \citealt{croton07} for a theoretical study of the assembly bias).}.
The red star forming galaxies that we find may be the optical counterparts
of such galaxies.  They are probably actively forming stars, as seen
from the [OII] emission, but contain large amounts of dust, which
reddens the rest frame optical colors.

The second finding - the relatively weak \hd\ line in group galaxies
- is unexpected. Star forming galaxies should have young stellar
populations, and we expect to observe strong \hd. 
The strength of the \hd\ line in field galaxies is within
expectations (the models can reproduce them as shown in the last section).
The strength of the \hd\ line in group galaxies is not.

The first point above may lead us to suggest that physical
processes that triggers a sharp truncation is working in groups
because starbursts can be induced by interactions,
which can cause a sharp decline in star formation rates after the bursts.
But, the second point may lead us to suggest the opposite because
a gradual truncation is needed to suppress \hd\ as shown
in the last section.  We argue in the following, however, that the second
point could also suggest a sharp truncation scenario.  To reach that
point, we first have to explore dust extinction.  It could be
possible that a selective extinction of young stars in star forming
regions absorb fluxes from early type stars and affect \hd.

\subsection{The effect of dust on the strength of \hd}

\begin{figure}
\centering
\includegraphics[width=9cm]{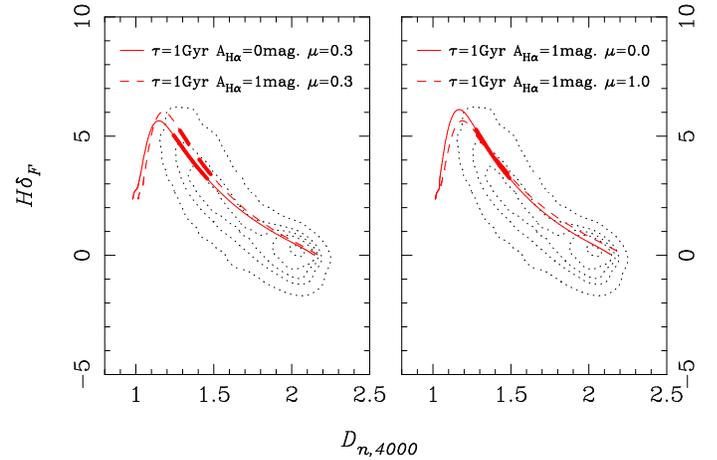}
\caption{
\hd\ plotted against \dn.
{\it Left panel:} The curves represent $\tau=1$\,Gyr star formation models with and without dust extinction.
The fraction of ambient interstellar extinction is $\mu=0.3$.
The models start from the left (0 Gyr) and evolve to the right (13 Gyr).
The thick part of the lines indicate the range of formation redshifts $2<z_f<5$ seen at $z=1.2$.
The contours show the distribution of galaxies from the Sloan survey
for comparison. 
The contours enclose 5, 25, 50, 75 and 95 percent of the galaxies population.
{\it Right panel:} $\tau=1$\,Gyr star formation models with an extinction of H$\alpha=1$ mag.
The solid and dashed curves are for models with $\mu=0$ (i.e., extinction
is entirely due to dust in star forming regions) and $\mu=1$ (i.e., extinction
is entirely due to ambient interstellar medium), respectively.
}
\label{fig:hd_d4000_dust}
\end{figure}

To quantify the effect of extinction on the \hd\ and \dn, we model
two levels of extinction: no extinction, i.e., A$_{\rm H\alpha}=0$\,mag,
and A$_{\rm H\alpha}=1$\,mag, which is typical for
local star forming galaxies \citep{kennicutt98}.  Since we do not know
the typical amount of extinction at this high redshift, we simply assume
that it is the same as $z=0$.  To model the selective extinction of young stellar
populations, we use the prescription described in
\citet{charlot00}, which is incorporated in the \citet{bruzual03} models.
Young stars ($<10^7$ yr) and older stars are affected by different
amounts of dust and, the way dust affects \hd\ and \dn\ is shown in
Fig. \ref{fig:hd_d4000_dust}.

A$_{\rm H\alpha}=1$\,mag of extinction makes \dn\ larger by $\sim0.1$
at most.  The \dn\ index is not strongly affected by dust because it
is measured in a relatively narrow wavelength window.  The \hdf\ index
is larger by $\sim0.5$ at its peak.  For the fractional contribution
of the ambient interstellar medium to the overall extinction, we used
$\mu=0.3$. This means, 70\% of the extinction comes from star forming
regions and it affects only newly born stars.  The rest of the 30\%
comes from interstellar medium and it affects all stars.  Young stars
($<10^7$ yr) born in star forming regions are therefore more obscured
than A-type stars, resulting in larger values for \hdf\ at its peak.
There is a large uncertainty in the time scale over which dust
can obscure young stars, but it is very unlikely that the dust hides
stars as long as $10^9$ yr to completely obscure A-type stars
\citep{blitz80}.

We can also explore how varying selective extinction affects both
\hdf\ and \dn.  The right panel in Fig. \ref{fig:hd_d4000_dust} shows
models with $\mu=0$ (i.e., the dust is confined to star forming
regions) and $\mu=1$ (i.e., the dust is distributed throughout the
interstellar medium of the entire galaxy).  \hdf\ is stronger at its
peak in the $\mu=0$ model for the same reason discussed above.  \dn\
is larger by $\sim0.1$ only at late times when old stars dominate the
flux.  While the effect of dust at $z=1.2$ might be somewhat larger
than we discuss here, it seems clear that dust alone is not able to
change \hdf\ and \dn\ significantly.  In particular, it cannot account
for the weak \hdf\ observed in group galaxies.  Dust can increase the
\hd\ absorption strengths, but cannot decrease them.

However, dust could affect the observed \hdf\ absorption in an
indirect way.  We used the \oii\ flux to correct for \hd\ emission
from HII regions and to obtain the stellar \hd\ absorption.  Dust
could affect this procedure, since the \oii\ emission is prone to
absorption by dust. The \oii\ fluxes we actually observe are
a lower limit of the unabsorbed \oii\ fluxes.  This will lead us to
underestimate the \hd\ emission, resulting in an underestimate to the
stellar \hd\ line strength.  In converting from a \oii\ flux to \ha\
flux and then to a \hd\ flux we assumed A$_{\rm H\alpha}=1$ mag. The
weak \hdf\ index is particularly noticeable for blue
($D_{n,4000}\sim1.2$) galaxies in groups.  \hdf\ in these galaxies is
$\sim3\AA$ smaller than \hdf\ in blue galaxies in the field.

If we assume that this difference is due to the star forming regions in
group galaxies being dustier than the star forming regions in field
galaxies, we find that A$_{\rm H\alpha}\sim5$\,mag of absorption in groups is
needed to bring the strength of \hdf\ in these galaxies up to level
observed in field galaxies.  This is a significant amount of
extinction, but it might not be too surprising given the result by
\citet{hicks02}, who obtained $E(B-V)\sim1.1$ ($A_{H\alpha}\sim3$)
from a handful of field galaxies at $0.8<z<1.6$.  Such a significant
amount of dust could strongly redden galaxies, but we do not observe
galaxies redder than the red sequence (Fig. \ref{fig:cmd_spec}).
\citet{conroy09} recently showed that colors of galaxies
are much less affected by dust if the dust distribution is more
clumpy.  Their result suggests that it is not very straightforward to
estimate dust contents of galaxies from broadband colors.  Emission
lines coming out of star forming regions may be a better tracer of
dust, but our extinction from \oii\ and \hd\ will probably not be
accurate enough to pin down the amount of dust in the groups.

We note that metallicity variations can also change \hdf,
but the effect is relatively small.
In \citet{bruzual03} models, super-solar ($Z=0.05$) and
subsolar ($Z=0.008$) models result in differences of $|\Delta{\rm
  H}\delta_F|\sim0.3\AA$ from the solar metallicity model.  \dn\ is
more strongly affected because it encompasses metal lines.
\citet{prochaska07} showed that \hdf\ measurements are affected by the
CN absorption on the redward side of \hd.  However, we observe a
strong offset in \hdf\ not only in red galaxies, but also in blue
galaxies for which we can safely ignore the effects of CN
\citep{prochaska07}.  Therefore, the observed \hdf\ offset is probably
not due to metallicity variations.

As discussed above, strong dust extinction in the groups can bring
the second point raised above into agreement with the first point.
They both could point us to a physical process that truncates
star formation on a short time scale.
An interesting point here is that the galaxies are dustier in
the groups, where the on-going build-up of the red sequence is happening.

\subsection{The physical process behind the build-up of the red sequence}

The strong extinction may be an important clue to the driving process
of the build-up of the red sequence.
The driving physical process of the red sequence needs to be (1)
effective in groups, (2) to trigger dusty starbursts, and (3) to
suppresses star formation afterwards. Galaxy-galaxy interactions, for example, should
occur most frequently in the group environment. These interactions can
trigger dusty starbursts (e.g., \citealt{bridge07}).
These large starbursts consume and expel a lot of the ingredients that
are needed for future star formation episodes. At the end of the
burst, there is little gas left for active star formation, which
results in a sharp decline in star formation rates.

The observed weak \hd\ absorptions may suggest that galaxy-galaxy
interactions in groups are the driving physical process of the
build-up of the red sequence and environment-dependent galaxy
evolution.  Galaxies exhaust their gas and stop their star formation
in poor groups before they finally merge with the cluster core.  This
scenario is supported by other observations of lower redshift clusters
\citep{tanaka06,tanaka07b,koyama08}.
\citet{elbaz07} and \citet{cooper08} showed that the average SFR
of galaxies increases with increasing galaxy density at $z\sim1$.
Although it is not straightforward to compare their results with ours,
a part of the increase might be driven by the possible dusty
starbursts in groups.
One may wonder why we do not observe weak \hdf\ in the field.
Starbursts may also happen in the field, but they may go unnoticed
because they are masked by the active underlying star formation.
It should be easier to detect them on top of relatively weak star formation.
Interactions alone may not be able to reproduce the down-sizing
phenomenon \citep{cowie96}, but massive galaxies may have experienced
more mergers in the past than low-mass galaxies and they might
have become red earlier.

While dust is one possible explanation, other explanations are possible.
For example, a weak star burst on top of old stellar population might
be able to reproduce the weak \hdf, although fine tuning of the relative
fraction of the old and new stellar populations would be needed.
We admit that it is not clear how to interpret Figs.  \ref{fig:hd_d4000} and
\ref{fig:hd_ssfr}, but it is clear that the {\it mode} of star formation is
different in the group and field environments, as indicated by the
systematic offset in \hd.  If it is because the star formation in
group galaxies is highly obscured, then we would expect to see clear
signatures of this in the near-infrared. For example, near infrared
spectroscopy of the \ha\ and \hb\ lines will
tell us the amount of dust in star forming regions.
High resolution imaging to
obtain morphology of galaxies is another way to test this
idea.  If group galaxies are undergoing starbursts triggered by
interactions, we should be able to see tidal features around some of
the galaxies.  In any case, it is clear that the {\it mode} of star
formation at $z\sim1.2$ depends on environment and that this may be an
important clue to address the origin of the environment-dependent
galaxy evolution.

\section{Conclusions}

We have carried out spectroscopic observations of the photometrically
selected groups embedded in the large-scale structure surrounding the
X-ray luminous galaxy cluster RDCS1252 at $z=1.24$. Three
groups contain galaxies at the cluster redshift. Two of them are
embedded in a filamentary structure that extends northwards from the
cluster core. The third group lies in a second structure that extends
eastwards from the cluster. It is likely that all three groups are
dynamically bound to the cluster. This is the first spectroscopic
confirmation of large scale structure surrounding such a high redshift
cluster.

We then quantified the spectral properties of galaxies at $z\sim1.2$
in the cluster, group, and field with the \dn, \ewoii, and \hdf\
indices.  In both the field and the groups, we find that there is a
significant population of \oii\ emitting galaxies that are notable for
their relatively red colors. The cluster core hosts few such
galaxies.  We also find that group galaxies have systematically
weaker \hdf\ than field galaxies.  We discussed the effects
that dust can have on \hdf\ measurements and suggested that a possible
way to interpret the weak \hdf\ is that the star forming regions
in group galaxies are more dusty than those in field galaxies.

We then suggested that galaxy-galaxy interactions lead to a higher
rate of dusty starbursts in group galaxies compared to the field.
Interactions should occur more frequently between galaxies in groups
than between field galaxies.  Starbursts are effective at consuming
and removing gas, which then leads to a sharp decrease in the star
formation rate once the burst ends, thus leading to more red
galaxies. We suggest that interactions are the driving physical
process in the build-up of the red sequence and environment-dependent
galaxy evolution.  This scenario may not be the only way to interpret
our results, but it is clear that the {\it mode} of star formation is
dependent on environment at $z\sim1.2$ as shown by the difference in
\hdf\ between group and field galaxies. Near-infrared observations to
measure \ha\ and high resolution imaging to obtain galaxy morphologies
can be used to test this scenario.

\begin{acknowledgements}

This study is based on observations obtained at the European Southern
Observatory using the ESO Very Large Telescope on Cerro Paranal
through ESO program 081.A-0759.
This study is also based on observations obtained at the Gemini Observatory
through Subaru-Gemini time exchange program GS-2006B-Q14.
The Gemini observatory is operated by the Association of Universities for
Research in Astronomy, Inc., under a cooperative agreement with the NSF on
behalf of the Gemini partnership:
the National Science Foundation (United States), the Science and Technology
Facilities Council (United Kingdom), the National Research Council (Canada),
CONICYT (Chile), the Australian Research Council (Australia), Ministério
da Ciência e Tecnologia (Brazil) and SECYT (Argentina).
In addition, we used data collected at Subaru Telescope,
which is operated by the National Astronomical Observatory of Japan,
and data taken at the United Kingdom Infrared Telescope, which
is operated by the Joint Astronomy Centre on behalf of the U.K.
Particle Physics and Astronomy Research Council.
The observations at the UKIRT 3.8-m telescope were supported by NAOJ.
Funding for the SDSS and SDSS-II has been provided by the Alfred
P. Sloan Foundation, the Participating Institutions, the National Science
Foundation, the U.S. Department of Energy, the National Aeronautics and
Space Administration, the Japanese Monbukagakusho, the Max Planck Society,
and the Higher Education Funding Council for England.
The SDSS Web Site is http://www.sdss.org/.
We thank Sune Toft for useful comments on the draft, and
Arjen van der Wel and Brad Holden for kindly providing us with their spectra.
We thank the anonymous referee for helpful comments, which improved the paper.
This work was financially supported in part by the Grant-in-Aid for
Scientific Research (Nos.\, 18684004 and 21340045) by the Japanese
Ministry of Education, Culture, Sports and Science.
AF has been partially supported trough NASA grant NNX08AD93G to UMBC.
\end{acknowledgements}

\bibliographystyle{aa}
\bibliography{12675}

\end{document}